\documentclass[structabstract]{aa}  
\usepackage{graphicx}
\usepackage{txfonts}
\usepackage{natbib}
\usepackage{tabularx}

\bibpunct[; ]{(}{)}{;}{a}{}{}
\begin{document}
   \title{Near-Infrared Properties of Metal-Poor Globular Clusters in the Galactic Bulge Direction
		\thanks{Based on observations carried out at the Canada-France-Hawaii
        Telescope, operated by the National Research Council of Canada,
        the Centre National de la Recherche Scientifique de France,
        and the University of Hawaii.
        This is a part of the survey for the Galactic bulge clusters using the CFHT,
        organized by the Korea Astronomy and Space Science Institute.}
		 }

   \author{S.-H. Chun\inst{1}, J.-W. Kim\inst{2}, I.-G. Shin\inst{1}, 
			C. Chung\inst{1}, D.-W. Lim\inst{1}, J.-H. Park\inst{3},
            H.-I. Kim\inst{3}, W. Han\inst{3}, \and Y.-J. Sohn\inst{1,3}
          }

   \institute{Department of Astronomy, Yonsei University, Seoul 120-749, Korea\\
              \email{sohnyj@yonsei.ac.kr, shchun@galaxy.yonsei.ac.kr}
         \and
             Institute for Computational Cosmology, Department of Physics, Durham University, 
			 South Road, Durham DH1 3LE, UK\
         \and
			Korea Astronomy and Space Science Institute, Daejeon 305-348, Korea     
			}

   \date{Received dd August 2009 / Accepted dd Monthber 2010}

 
  \abstract
   {}
    {$J$, $H$, and $K^{'}$ images obtained from the near-infrared imager CFHTIR on the 
	Canada-France-Hawaii Telescope are used to derive the morphological parameters of 
	the red giant branch (RGB) in the near-infrared color-magnitude diagrams 
	for 12 metal-poor globular clusters in the Galactic bulge direction.
	Using the compiled data set of the RGB parameters for the observed 12 clusters, 
	in addition to the previously studied 5 clusters,
	 we discuss the properties of the RGB morphology for the clusters and compare them with 
    the calibration relations for the metal-rich bulge clusters and the metal-poor halo clusters.
	}
    {The photometric RGB shape indices such as colors at fixed magnitudes of 
	$M_K=M_H=(-5.5, -5, -4,$ and $-3)$, magnitudes at fixed colors of $(J-K)_o = (J-H)_o = 0.7$,
	and the RGB slope are measured from the fiducial normal points
	defined in the near-infrared color-magnitude diagrams for each cluster.
	The magnitudes of RGB bump and tip are also estimated from
	the differential and cumulative luminosity functions of the selected RGB stars.
	The derived RGB parameters have been used to examine the overall behaviors of the RGB morphology
	as a function of cluster metallicity.}
  	{The correlations between the near-infrared photometric RGB shape indices and the cluster metallicity
	for the programme clusters compare favorably with the previous observational calibration relations 
	for metal-rich clusters in the Galactic bulge and the metal-poor halo clusters. 
	The observed near-infrared magnitudes of the RGB bump and tip for the investigated clusters
	are also in accordance with the previous calibration relations for the Galactic bulge clusters.	
	}
   {}
   
   \authorrunning{Chun et al.}
   \titlerunning{Metal-poor globular clusters in the Galactic bulge direction}
   \maketitle
%

\section{Introduction}

\begin{table}
\caption{Observational log of the target clusters}      
\label{tbl1}
\centering
\begin{tabular}{c c c c c c}
\hline\hline
Target & Filter & Exp. time (sec) & FWHM$(^{''})$ & Year \\    
\hline
  NGC 6333 & $J$     & 4$\times$1, 8$\times$30 & 0.60, 0.67 & 2002\\      
           & $H$     & 4$\times$1, 8$\times$30 & 0.61, 0.58 &\\
           & $K^{'}$ & 4$\times$1, 8$\times$30 & 0.58, 0.61 &\\
  NGC 6626 & $J$     & 4$\times$1, 8$\times$30 & 0.67, 0.62 & 2002\\      
           & $H$     & 4$\times$1, 8$\times$30 & 0.57, 0.65 &\\
           & $K^{'}$ & 4$\times$1, 8$\times$30 & 0.62, 0.65 &\\
  NGC 6235 & $J$     & 4$\times$2, 8$\times$30 & 0.73, 0.74 & 2003\\      
           & $H$     & 4$\times$2, 8$\times$30 & 0.77, 0.81 &\\
           & $K^{'}$ & 4$\times$2, 8$\times$30 & 0.72, 0.77 &\\
  NGC 6266 & $J$     & 4$\times$2, 8$\times$30 & 0.82, 0.93 & 2003\\      
           & $H$     & 4$\times$2, 8$\times$30 & 0.73, 0.88 &\\
           & $K^{'}$ & 4$\times$2, 8$\times$30 & 0.77, 0.92 &\\
  NGC 6273 & $J$     & 4$\times$2, 8$\times$30 & 0.86, 0.87 & 2003\\      
           & $H$     & 4$\times$2, 8$\times$30 & 0.83, 0.74 &\\
           & $K^{'}$ & 4$\times$2, 8$\times$30 & 0.83, 0.74 &\\
  NGC 6287 & $J$     & 4$\times$2, 8$\times$30 & 0.61, 0.76 & 2003\\      
           & $H$     & 4$\times$2, 8$\times$30 & 0.62, 0.73 &\\
           & $K^{'}$ & 4$\times$2, 8$\times$30 & 0.65, 0.73 &\\
  NGC 6293 & $J$     & 4$\times$2, 8$\times$30 & 0.89, 1.18 & 2003\\      
           & $H$     & 4$\times$2, 8$\times$30 & 0.95, 0.87 &\\
           & $K^{'}$ & 4$\times$2, 8$\times$30 & 0.66, 0.73 &\\
  NGC 6325 & $J$     & 4$\times$2, 8$\times$30 & 0.73, 0.77 & 2003\\      
           & $H$     & 4$\times$2, 8$\times$30 & 0.65, 0.78 &\\
           & $K^{'}$ & 4$\times$2, 8$\times$30 & 0.68, 0.73 &\\
  NGC 6355 & $J$     & 4$\times$2, 8$\times$30 & 0.74, 0.80 & 2003\\      
           & $H$     & 4$\times$2, 8$\times$30 & 0.73, 0.87 &\\
           & $K^{'}$ & 4$\times$2, 8$\times$30 & 0.73, 0.83 &\\
  NGC 6401 & $J$     & 4$\times$2, 8$\times$30 & 0.66, 0.68 & 2003\\      
           & $H$     & 4$\times$2, 8$\times$30 & 0.61, 0.72 &\\
           & $K^{'}$ & 4$\times$2, 8$\times$30 & 0.60, 0.72 &\\
  NGC 6558 & $J$     & 4$\times$2, 8$\times$30 & 0.95, 0.99 & 2003\\      
           & $H$     & 4$\times$2, 8$\times$30 & 0.80, 0.99 &\\
           & $K^{'}$ & 4$\times$2, 8$\times$30 & 0.73, 0.89 &\\
  Terzan 4 & $J$     & 4$\times$1, 8$\times$30 & 0.68, 0.75 & 2003\\      
           & $H$     & 4$\times$1, 8$\times$30 & 0.65, 0.73 &\\
           & $K^{'}$ & 4$\times$1, 8$\times$30 & 0.80, 0.72 &\\
\hline
\end{tabular}
\end{table}

The current view of the Galaxy formation is mainly focused
on the hierarchical merging paradigm in the cold dark matter cosmology.
Globular clusters, as tracers of the early formation and
the current structure of the Galaxy, play a key role in studies of the paradigm,
because they are present from the central bulge to the outer halo with various metallicities.
Particularly, the Galactic bulge harbors a globular cluster population 
with a broad metallicity distribution that extends from about twice solar
to less than one-tenth solar abundance~\citep{Ort99}, 
while most field stars in the bulge have near-solar metallicity~\citep{Mcw94,Zoc03}. 
The metal-rich globular clusters in the the Galactic bulge 
share the kinematics, spatial distribution,
and composition of the bulge field stars~\citep[e.g.,][and references therin]{Min08}.
This indicates that metal-rich globular clusters are associated with the Galactic bulge
recognized as the dominant proto-Galactic building block~\citep[e.g.,][]{Cot00}.
On the other hand, the origin of the metal-poor globular clusters in the Galactic bulge direction
is still a subject of debate since accurate measurements of kinematics and 
high resolution chemical abundances are lacking.
In the hierarchical model of the Galaxy formation, however, 
old metal-poor field stars in the bulge form via merging and accretion events 
in the early Universe~\citep{Nak03}.
In this sense, the metal-poor clusters currently located in the central region
of the Galaxy might be the oldest objects which did not form originally in the Galactic bulge.
Thus, the metal-poor clusters in the bulge region can play a key role to understand
the early epoch of the formation of the Galactic bulge.

In the Milky Way, about 150 globular clusters are listed in the database of ~\citet{Har96}, 
which was revised in 2003. Recently, new faint clusters and cluster candidates have also been found
~\citep[e.g.,][]{Car05,Kob05,Wil05,Fro07}.
Out of 43 globular clusters located within 3 kpc of the Galactic center~\citep{Har96},
22 are metal-poor ([Fe/H] $<-1.0$) and 21 are metal-rich ([Fe/H] $>-1.0$).
Recently, ~\citet{Val07} presented near-infrared color-magnitude diagrams (CMDs) and physical parameters
for a sample of 24 globular clusters toward the Galactic bulge and located 
within $|b|\leqq10^{\circ}$ and $|l|\leqq20^{\circ}$.
They discussed the near-infrared properties of the red giant branch (RGB) for 
12 observed clusters, in addition to those previously studied by their group
~\citep[e.g.,][]{Fer00,Val04a,Val04b,Val04c,Val05,Ori05}.
We note, however, that their sample of the clusters have high priority to the 
metal-rich population, i.e., 17 out of the 24 are relatively metal-rich
with [Fe/H] $>-1.0$, taking into account of a bulge origin for the metal-rich globular clusters.

In our research, we have focused on obtaining a moderately deep homogeneous photometric data set 
in the near-infrared regime for the metal-poor clusters in the bulge direction.
Near-infrared photometry offers advantages for a study of the cool population of the RGB stars 
in the Galactic globular clusters,
because of its high sensitivity to low temperature.
In addition, high extinctions toward the bulge can be reduced by observing
the near-infrared wavelengths, as the extinction in the $K$ band is
only $\sim10$ percent of that in the $V$ band~\citep{Rie85}.
Using these bases, ~\citet{Kim06} presented the morphological properties of the RGB
in the near-infrared CMDs for five metal-poor clusters
of the Galactic bulge (NGC 6541, NGC 6642, NGC 6681, NGC 6717, and NGC 6723) 
and also for three halo clusters.

In this paper, we report new results of the near-infrared photometry for 12 metal-poor clusters
and present a homogeneous photometric data set of the 
RGB morphology for 17 globular clusters, covering $\sim75\%$ of the total 22 metal-poor globular 
clusters in the Galactic bulge direction.
The results of the RGB morphology for the programme clusters are compared with
the previously published calibrations of ~\citet{Val04a,Val04b,Val07}.
The observations, procedures for data reduction, and phtometric measurements are described in Sect. 2.
In Sect. 3, we describe the near-infrared CMDs and the fiducial normal points of target clusters.
In Sect. 4, the morphological properties of CMDs such as RGB shape feature, RGB bump, and RGB tip are presented.
Finally, the results are discussed and summarized in Sect. 5.

\section{OBSERVATIONS, DATA REDUCTION, AND PHOTOMETRIC MEASUREMENTS}
\begin{figure*}
\centering
\includegraphics[width=0.94\textwidth]{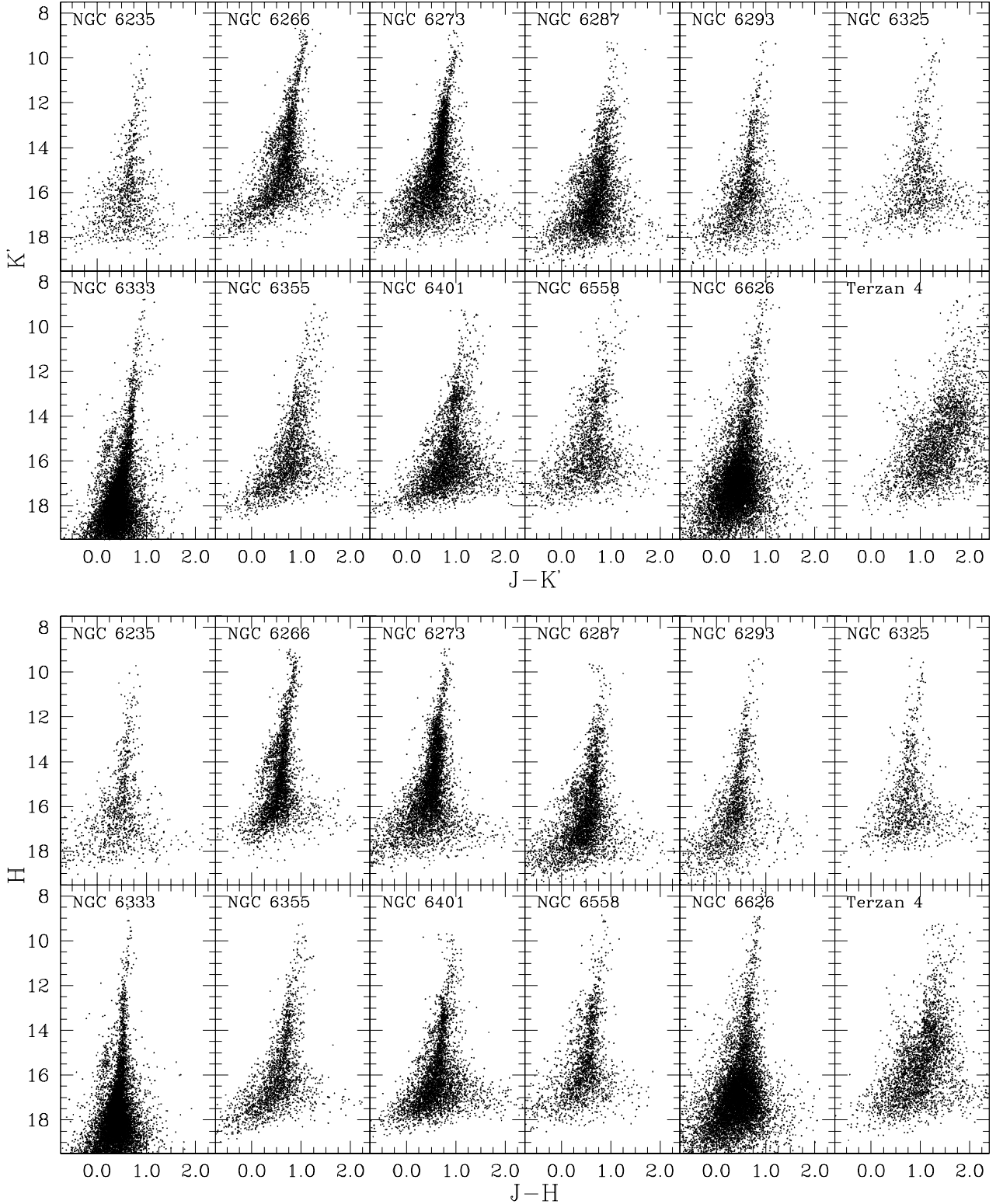}
\caption{The upper and lower panels are $(J-K^{'},K^{'})$ and $(J-H,H)$ CMDs of
the observed 12 clusters.}
\label{Fig1}
\end{figure*}

\begin{figure*}
\centering
\includegraphics[width=0.94\textwidth]{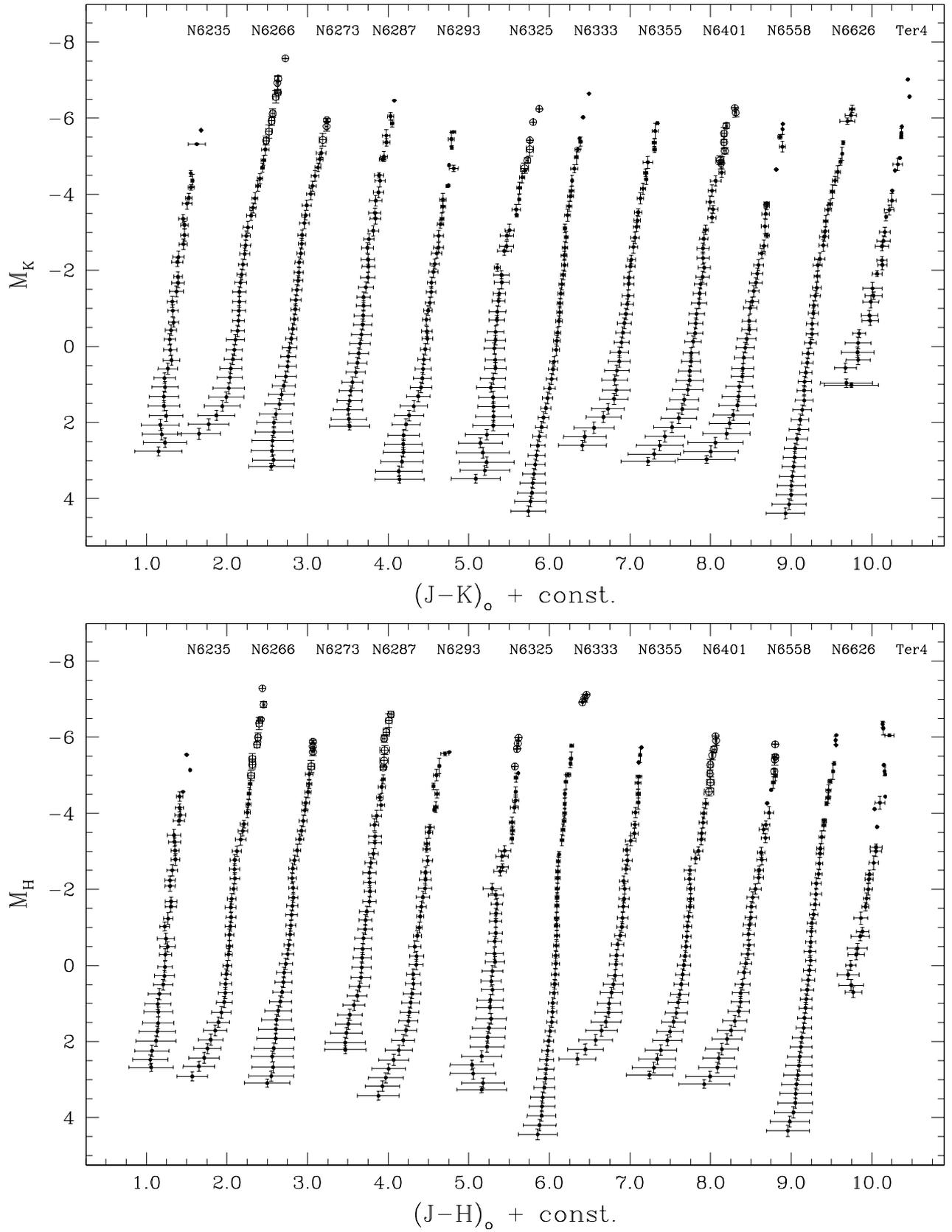}
\caption{Fiducial normal points of target clusters in $(J-K)_o-M_K$ ({\it upper}) and $(J-H)_o-M_H$ ({\it lower}) planes.
               For clarity, the colors are given zero-point offsets; from left to right, these are
               const. $=0.8, 1.6, 2.4, 3.2, 4.0, 4.8, 5.6, 6.4, 7.2, 8.0, 8.8$ and $9.6$ magnitudes.
               Open circles indicate fiducial normal points determined from the 2MASS data.}
\label{Fig2}
\end{figure*}

\begin{table*}
\caption{Metallicity, distance modulus, reddening, and
extinction values of the observed 12 globular clusters in the Galactic bulge.}
\label{tbl2}
\centering
\begin{tabular}{c c c c c c c c} 
\hline\hline
Target & [Fe/H]$_{CG97}$ & [M/H] & $\mu_o$ & $E(B-V)$ & $A_J$ & $A_H$ & $A_K$ \\
\hline
 NGC 6235  &  -1.17 & -0.97 & 15.05 & 0.36 & 0.325 & 0.207 & 0.132 \\
 NGC 6266  &  -1.08 & -0.87 & 14.26 & 0.47 & 0.424 & 0.271 & 0.172 \\
 NGC 6273  &  -1.45 & -1.24 & 14.58 & 0.41 & 0.370 & 0.236 & 0.150 \\
 NGC 6287  &  -1.90 & -1.68 & 15.61 & 0.60 & 0.541 & 0.346 & 0.220 \\
 NGC 6293  &  -1.73 & -1.53 & 14.79 & 0.41 & 0.370 & 0.236 & 0.150 \\
 NGC 6325  &  -1.21 & -0.99 & 14.10 & 0.89 & 0.803 & 0.513 & 0.327 \\
 NGC 6333  &  -1.56 & -1.36 & 14.67 & 0.38 & 0.343 & 0.219 & 0.139 \\
 NGC 6355  &  -1.26 & -1.07 & 14.60 & 0.75 & 0.677 & 0.432 & 0.275 \\
 NGC 6401  &  -0.97 & -0.74 & 14.61 & 0.72 & 0.649 & 0.415 & 0.264 \\
 NGC 6558  &  -1.21 & -0.99 & 14.30 & 0.44 & 0.397 & 0.253 & 0.161 \\
 NGC 6626  &  -1.21 & -0.99 & 13.60 & 0.40 & 0.361 & 0.230 & 0.147 \\
 Terzan 4  &  -1.62 & -1.41 & 15.10 & 2.06 & 1.858 & 1.187 & 0.756 \\
\hline
\end{tabular}
\end{table*}

Observations were obtained during the nights of June 1, 2002, and April 20-21, 2003.
Using the CFHTIR imager mounted on the f/8 Cassegrain focus of the Canada-France-Hawaii telescope (CFHT),
the fields centered on each cluster were observed with $J$, $H$ and $K^{'}$ filters.
The CFHTIR contains a $1024\times1024$ Hg:Cd:Te array.
Its angular scale is $0.211^{''}$/pixel, so that each image covers a total field-of-view
of $3.6^{'}\times3.6^{'}$.
The observations were split into short and long exposures for each filter
in order to optimize the photometry of bright and faint stars.
The images by short and long exposures are
combinations of four 1-second or 2-second exposures,
and of eight 30-second exposures, respectively.
A four-points square dither pattern was used to identify and reject bad pixels and cosmic rays in the observed images.
In both run the UKIRT standard stars and M13 were also observed for a photometric standardization.
The summary of observations for the target clusters is presented in Table~\ref{tbl1}.

Calibration frames of darks, flats, and blank sky backgrounds were also obtained during the runs.
Dark frames were recorded at the beginning and the end of each run.
Dome flats were made by subtracting exposures of the dome white spot taken with the lamps off
from those taken with the lamps on. Thermal emission patterns were constructed
by combining flat-fielded images of blank sky regions.

The process of data reduction consists of subtracting a dark frame,
dividing by the normalized flat image for each filter, and
subtracting the thermal signature and
the sky background level estimated by the mode of pixel intensity distribution.
Then, the processed images were combined for each exposure after aligning the dither offsets.
The seeing conditions of the reduced images range between $0^{''}.6\sim0^{''}.9$.

The brightness of stars in the clusters was measured with
the point-spread function fitting routine DAOPHOT II/ALLSTAR~\citep{Stet87,Stet88}.
The brightness of stars around the RGB tip was measured only in short-exposure images because of
saturation in long-exposure images, while faint stars were detected in only long-exposure images.
For stars detected in both short and long exposures, measurements with smaller photometric error
were assigned to the brightness.
To avoid false detection, only stars detected in all filters with detection errors of
less than 0.2 mag were considered for the photometric analysis.
The photometric calibration equations obtained from UKIRT standard stars
were then applied to the magnitudes of the stars on the target clusters.
Standardizations were also double-checked 
in direct star-to-star comparison with the photometry of bright stars
in M13 of ~\citet{Kim06}.
As shown in ~\citet{Kim06}, there are only small photometric offsets, 
$\vartriangle K=0.03\pm0.01$ and $\vartriangle(J-K)=0.04\pm0.01$,
between the photometric data with the UKIRT system
and those with the 2MASS system of ~\citet{Val04c}. 
Note that the offsets will become negligible after the transformation
of the fiducial normal points for the observed near-infrared CMD 
into the 2MASS photometric system (see Sec. 3).
Here, we also note that the measured photometric data in the south-west quarter part of the images
for the runs of 2002, and those in the south-east quarter part of the images for the runs of 2003,
were not used for the subsequent photometric analyses,
because of possible readout anomalies of the CFHTIR imager during the runs.

\section{COLOR-MAGNITUDE DIAGRAMS AND FIDUCIAL NORMAL POINTS}
Figure~\ref{Fig1} shows $(J-K^{'},K^{'})$ and $(J-H,H)$ CMDs of the resolved stars in the
observed area for the clusters investigated in this study.
As can be seen, all of the observations are deep enough to reach the base of the RGB
at $\Delta K^{'}\sim\Delta H\approx8$ mag fainter than the RGB tip.
As we expected in the near-infrared CMDs for metal-poor globular clusters,
the horizontal branch sequences are aslant compared to the RGB sequences.
Scattering in the near-infrared CMDs of the target clusters might be due to 
both photometric errors and contamination by foreground field stars toward the Galactic bulge.
Apparently in the case of the highly reddened tiny cluster Terzan 4, for which
~\citet{Bon08} derived the tidal radius of $7.6^{'}\pm1.3^{'}$
and the concentration parameter of $c=0.9\pm0.2$ from the 2MASS images,
the CMDs contain significant noise owing to a strong field star contamination
in the observed field.

\begin{table*}
\caption{The RGB colors of the observed bulge clusters at different magnitudes.}
\label{tbl3}
\centering
\begin{tabular}{c c c c c c c c c}
\hline\hline
Target & $(J-K)_o^{-5.5}$ & $(J-K)_o^{-5}$ & $(J-K)_o^{-4}$ &
$(J-K)_o^{-3}$ & $(J-H)_o^{-5.5}$ & $(J-H)_o^{-5}$ & $(J-H)_o^{-4}$ &
$(J-H)_o^{-3}$ \\
\hline
\multicolumn{9}{c}{This paper} \\
\hline
NGC 6235    & 0.853$\pm$0.04 & 0.794$\pm$0.04 & 0.716$\pm$0.04 & 0.659$\pm$0.05 & 0.697$\pm$0.05 & 0.676$\pm$0.05 &
0.624$\pm$0.05 & 0.550$\pm$0.06\\
NGC 6266    & 0.908$\pm$0.03 & 0.864$\pm$0.03 & 0.761$\pm$0.04 & 0.657$\pm$0.06 & 0.737$\pm$0.03 & 0.699$\pm$0.03 &
0.645$\pm$0.04 & 0.531$\pm$0.06\\
NGC 6273    & 0.812$\pm$0.04 & 0.761$\pm$0.04 & 0.635$\pm$0.05 & 0.543$\pm$0.05 & 0.658$\pm$0.04 & 0.634$\pm$0.04 &
0.563$\pm$0.04 & 0.482$\pm$0.05\\
NGC 6287    & 0.794$\pm$0.05 & 0.755$\pm$0.05 & 0.677$\pm$0.05 & 0.597$\pm$0.08 & 0.649$\pm$0.05 & 0.622$\pm$0.05 &
0.571$\pm$0.05 & 0.517$\pm$0.07\\
NGC 6293    & 0.793$\pm$0.02 & 0.766$\pm$0.02 & 0.716$\pm$0.04 & 0.643$\pm$0.05 & 0.677$\pm$0.03 & 0.608$\pm$0.03 &
0.536$\pm$0.03 & 0.486$\pm$0.05\\
NGC 6325    & 0.962$\pm$0.02 & 0.938$\pm$0.02 & 0.809$\pm$0.04 & 0.682$\pm$0.05 & 0.788$\pm$0.03 & 0.754$\pm$0.03 &
0.665$\pm$0.05 & 0.590$\pm$0.06\\
NGC 6333    & 0.778$\pm$0.02 & 0.741$\pm$0.02 & 0.665$\pm$0.03 & 0.611$\pm$0.03 & 0.669$\pm$0.02 & 0.633$\pm$0.02 &
0.588$\pm$0.02 & 0.527$\pm$0.02\\
NGC 6355    & 0.912$\pm$0.04 & 0.864$\pm$0.04 & 0.748$\pm$0.04 & 0.664$\pm$0.05 & 0.725$\pm$0.04 & 0.693$\pm$0.04 &
0.642$\pm$0.04 & 0.580$\pm$0.05\\
NGC 6401    & 0.985$\pm$0.04 & 0.935$\pm$0.04 & 0.851$\pm$0.04 & 0.758$\pm$0.06 & 0.825$\pm$0.03 & 0.790$\pm$0.03 &
0.723$\pm$0.03 & 0.641$\pm$0.04\\
NGC 6558    & 0.876$\pm$0.05 & 0.841$\pm$0.05 & 0.760$\pm$0.05 & 0.671$\pm$0.05 & 0.802$\pm$0.04 & 0.777$\pm$0.04 &
0.706$\pm$0.04 & 0.626$\pm$0.04\\
NGC 6626    & 0.886$\pm$0.04 & 0.826$\pm$0.04 & 0.707$\pm$0.04 & 0.618$\pm$0.03 & 0.744$\pm$0.03 & 0.705$\pm$0.03 &
0.623$\pm$0.03 & 0.564$\pm$0.04\\
Terzan 4    & 0.769$\pm$0.03 & 0.727$\pm$0.03 & 0.650$\pm$0.03 & 0.579$\pm$0.03 & 0.650$\pm$0.05 & 0.616$\pm$0.05 &
0.547$\pm$0.05 & 0.477$\pm$0.05\\
\hline
 \multicolumn{9}{c}{~\citet{Kim06}}  \\
\hline
NGC 6541    & 0.824$\pm$0.02 & 0.789$\pm$0.02 & 0.687$\pm$0.02 & 0.595$\pm$0.02 & 0.695$\pm$0.02 & 0.663$\pm$0.02 &
0.584$\pm$0.02 & 0.511$\pm$0.03\\
NGC 6642    & 0.924$\pm$0.03 & 0.843$\pm$0.03 & 0.720$\pm$0.03 & 0.632$\pm$0.04 & 0.816$\pm$0.03 & 0.759$\pm$0.03 &
0.659$\pm$0.03 & 0.578$\pm$0.04\\
NGC 6681    & 0.826$\pm$0.02 & 0.778$\pm$0.02 & 0.686$\pm$0.02 & 0.628$\pm$0.02 & 0.680$\pm$0.03 & 0.644$\pm$0.03 &
0.584$\pm$0.03 & 0.537$\pm$0.03\\
NGC 6717    & 0.968$\pm$0.03 & 0.887$\pm$0.03 & 0.777$\pm$0.03 & 0.714$\pm$0.03 & 0.820$\pm$0.02 & 0.796$\pm$0.02 &
0.727$\pm$0.02 & 0.650$\pm$0.02\\
NGC 6723    & 0.940$\pm$0.02 & 0.899$\pm$0.02 & 0.787$\pm$0.03 & 0.696$\pm$0.03 & 0.751$\pm$0.03 & 0.717$\pm$0.03 &
0.665$\pm$0.04 & 0.580$\pm$0.03\\
\hline
\end{tabular}
\end{table*}

To examine the relationship between the RGB morphological parameters in CMDs of the absolute plane
and cluster's metallicity, the values of metallicity, reddening, and distance modulus
were estimated for each cluster using the method adopted in ~\citet{Kim06}.
Metallicities for target clusters are used in the~\citet{Carr97} scale, [Fe/H]$_{CG97}$,
to directly compare the photometric properties of the measured RGB morphology
with the results presented in~\citet{Val04a,Val07}.
Metallicities [Fe/H]$_{CG97}$ of two clusters NGC 6266 and NGC 6333 were adopted from~\citet{Fer99}.
For the other clusters, we obtained [Fe/H]$_{CG97}$ by transforming the data given in~\citet{Zin85}
into the scale of~\citet{Carr97} as per~\citet{Val04a}.
Note that we assigned [Fe/H]$_{CG97}$ $=-1.62\pm0.08$ to Terzan 4,
for which~\citet{Ste04} measured the metallicity of 7 stars in the cluster.
We also estimated global metallicities [M/H] of the target clusters
by using the equation for the $\alpha$ elements enhanced theoretical evolutionary sequence~\citep{Sal93},
i.e., [M/H]$=$[Fe/H]$_{CG97}+log(0.638f_\alpha+0.362)$ with $f_{\alpha}=10^{0.30}$,
where $f_\alpha$ is the enhancement factor of the $\alpha$ elements.
The determined metallicity values of [Fe/H]$_{CG97}$ and [M/H] for each cluster are listed in Table~\ref{tbl2}.

Distance moduli of two clusters NGC 6266 and NGC 6333 were adopted from~\citet{Fer99},
in which a new methodology is presented to derive distance moduli of globular clusters
by matching the observed visual magnitude of the zero-age horizontal branch ($V_{ZAHB}$) and
the theoretical synthetic horizontal branch (HB) models.
For the other ten clusters,
a similar procedure to that of~\citet{Fer99} was applied to determine
the distance moduli from the synthetic and observed ZAHB levels.
Synthetic HBs for each cluster with different metallicities were generated by 
the method used in ~\citet{Lee94} with the HB evolutionary tracks of~\citet{Yi97}.
Details of the generated synthetic HBs with various metallicities are described in ~\citet{Kim06}.
The synthetic HBs in absolute plane were transformed into the observed HBs in the CMDs
of the target clusters from~\citet{Rich98} for NGC 6558, ~\citet{Ort97} for Terzan 4,
and~\citet{Pio02} for the other eight clusters.
The extinction correction was calculated by using the latest compilation of
$E(B-V)$ in ~\citet{Har96} and by applying the reddening ratios of~\citet{Sch98}.
The distance modulus for each cluster was then estimated by measuring
the ZAHB levels in the synthetic and observed CMDs of HB stars,
taking into account the extinction values for each cluster.
Note that we determined the reddening $E(B-V) = 2.06$ for the highly reddened cluster Terzan 4
from the synthetic and observed CMDs of HB stars. This seems to be
slightly smaller than $E(B-V) = 2.35$ in ~\citet{Har96} and $E(B-V) = 2.31$ in~\citet{Ort97},
but comparable to the reddening value $E(B-V)=2.05$ of ~\citet{Val10}.
The determined distance moduli $\mu_o$ for the target clusters are listed in Table~\ref{tbl2}
with reddening $E(B-V)$ and extinction values $A_J$, $A_H$, and $A_K$ in the near-infrared wavelengths.

Prior to the derivation of the morphological parameters of the RGB sequence,
we obtained the RGB fiducial normal points for the near-infrared CMDs of the sample clusters,
following the same strategy as in~\citet{Kim06}.
As shown in Figure~\ref{Fig1}, the CMD of Terzan 4 shows a significant field star 
contamination. 
In order to minimize field star contamination of the tiny cluster Terzan 4,
we determined the fiducial normal points of the RGB 
with stars only within $16^{''}$ of the cluster center.
Note that ~\citet{Val10} derived the RGB ridge line of Terzan 4 
using stars within $40^{''}$ of the cluster center to derive the morphological parameters.
For the other clusters, the resolved stars within $30^{''}$ from the cluster center were
used to construct the fiducial normal points of the RGB.
We first determined the mean magnitude and color in the 0.25 mag bin of the CMDs,
excluding asymptotic giant branch stars, slanted HB stars, and highly scattered foreground stars.
Subsequently, we rejected stars with colors larger than $\pm2\sigma$ of the mean,
and the mean values of the magnitude and color were calculated again in the assigned magnitude bin.
The procedure with a $2\sigma$ rejection criterion was repeated until the mean
values of the magnitude and color are stable at constant values.
This iterative process statistically removed the asymptotic giant branch stars, HB stars, and field stars
from the RGB stars in the obtained near-infrared CMDs for
the central region of the target clusters.
Then, the cluster reddening and distance were used to convert the determined
fiducial normal points into the absolute plane.
Finally, the color and magnitude of the fiducial normal points in the UKIRT system
were transformed into the 2MASS system by using equations (37)-(39) from ~\citet{Carp01}
to compare the results directly with those of ~\citet{Val04a,Val07}.
Figure~\ref{Fig2} shows the fiducial normal points in $(J-K)_o-M_K$ and $(J-H)_o-M_H$ planes for the target clusters.
In the case of bright stars saturated around the RGB tip,
we estimated the fiducial normal points from the 2MASS catalog data of the area
observed in this study, and those are represented in Figure~\ref{Fig2} by open circles.

\section{MORPHOLOGY OF THE NEAR-INFRARED CMDS}

In this section, we present and discuss the morphological properties of
the near-infrared CMDs for the programme clusters.
The near-infrared RGB morphology for each cluster are characterized by
parameters of the RGB location in colors at fixed magnitudes and in magnitudes at fixed colors,
the slopes of the RGB, and the absolute magnitudes of the RGB bumps and tips.
The RGB parameters for 12 clusters in this paper which together with the 5 clusters in~\citet{Kim06}
have been used to examine the overall behaviors of the RGB
morphology in the near-infrared CMDs as a function of cluster metallicity 
for the metal-poor globular clusters in the Galactic bulge direction.
The results were compared with the previous observational calibrations
of~\citet{Val04a,Val07} and
the theoretical predictions of the Yonsei-Yale isochrones~\citep{Kim02,Yi03}.

\subsection{The RGB Shape}

To characterize the overall behaviors of the RGB morphology in the near-infrared and optical CMDs
of globular clusters,~\citet{Fer00} defined a new set of
photometric indices for the RGB location, i.e., colors at fixed magnitudes and magnitudes at fixed colors.
In a similar fashion,~\citet{Kim06} measured the photometric color and magnitude indices
of the RGB morphology for five metal-poor globular clusters in the bulge direction, and
compared the results with calibrations of the RGB morphology
for 28 bulge clusters from ~\citet{Val04a,Val05}.
The representative morphological parameters of the RGB include
(1) $(J-K)_o$ and $(J-H)_o$ colors at four fixed absolute magnitude levels of $M_K=M_H=(-5.5, -5, -4,$ and $-3)$,
(2) the absolute magnitudes of $M_K$ and $M_H$ at fixed colors of $(J-K)_o=(J-H)_o=0.7$, and
(3) the slope in the $(J-K, K)$ color-magnitude plane.

In the present study, we also measured the same parameters for the observed clusters.
To derive the RGB location parameters in color and in magnitude,
we applied a second- or third- order polynomial fit to adjacent $\gtrsim10$ fiducial normal points
of CMDs in Figure~\ref{Fig2} at the given magnitude and color.
The RGB slope has usually been determined by fitting an equation of the form
$J-K=aK+b$ to the upper part of the RGB in the $(J-K, K)$ CMD,
where the RGB morphology is less curved than in the other lower faint ranges.
In the same manner as ~\citet{Kim06}, the fiducial normal points in a 
magnitude range between 0 and 5 mag fainter than the brightest point 
were used to determine the RGB slope.
Table~\ref{tbl3} lists the determined $(J-K)_o$ and $(J-H)_o$ colors at different magnitude levels.
Furthermore, the absolute magnitudes $M_K$ at $(J-K)_o=0.7$ and $M_H$ at $(J-H)_o=0.7$,
and the RGB slope are presented in Table~\ref{tbl4}.
In Table 3 and Table 4, we also list previously studied RGB shape parameters by our group
~\citep[i.e.,][]{Kim06} for 5 metal-poor clusters in the bulge direction.

\begin{table}
\caption{The RGB magnitudes at different colors, and the RGB slopes for the observed bulge clusters.}
\label{tbl4}
\centering
\begin{tabular}{c c c c}
\hline\hline
Target & $M_K^{(J-K)_o=0.7}$ & $M_H^{(J-H)_o=0.7}$ & $RGB_{slope}$ \\
\hline
\multicolumn{4}{c}{This paper} \\
\hline
NGC 6235 & -3.55$\pm$0.64 & -5.56$\pm$1.11 & -0.068$\pm$0.006\\
NGC 6266 & -3.49$\pm$0.57 & -4.90$\pm$0.45 & -0.098$\pm$0.008\\
NGC 6273 & -4.60$\pm$0.36 &     ...        & -0.084$\pm$0.009\\
NGC 6287 & -4.45$\pm$1.11 & -6.17$\pm$0.76 & -0.069$\pm$0.008\\
NGC 6293 & -4.03$\pm$0.42 & -5.63$\pm$0.16 & -0.066$\pm$0.004\\
NGC 6325 & -3.10$\pm$0.51 & -4.30$\pm$0.35 & -0.106$\pm$0.010\\
NGC 6333 & -4.55$\pm$0.46 & -6.06$\pm$0.26 & -0.064$\pm$0.003\\
NGC 6355 & -3.48$\pm$0.64 & -5.14$\pm$0.59 & -0.078$\pm$0.008\\
NGC 6401 & -2.09$\pm$0.97 & -3.72$\pm$0.47 & -0.087$\pm$0.010\\
NGC 6558 & -3.22$\pm$0.58 & -3.73$\pm$0.63 & -0.083$\pm$0.004\\
NGC 6626 & -3.93$\pm$0.53 & -5.01$\pm$0.47 & -0.095$\pm$0.005\\
Terzan 4 & -4.68$\pm$0.37 & -6.28$\pm$0.79 & -0.075$\pm$0.005\\
\hline
\multicolumn{4}{c}{\citet{Kim06}} \\
\hline
NGC 6541 & -4.14$\pm$0.23 & -5.56$\pm$0.34 & -0.072$\pm$0.003\\
NGC 6642 & -3.71$\pm$0.28 & -4.51$\pm$0.37 & -0.104$\pm$0.006\\
NGC 6681 & -4.10$\pm$0.25 & -5.94$\pm$0.24 & -0.075$\pm$0.003\\
NGC 6717 & -2.73$\pm$0.33 & -3.66$\pm$0.25 & -0.077$\pm$0.004\\
NGC 6723 & -3.05$\pm$0.37 & -4.87$\pm$0.50 & -0.082$\pm$0.003\\
\hline
\end{tabular}
\end{table}

   \begin{figure}
   \centering
   \includegraphics[width=0.45\textwidth]{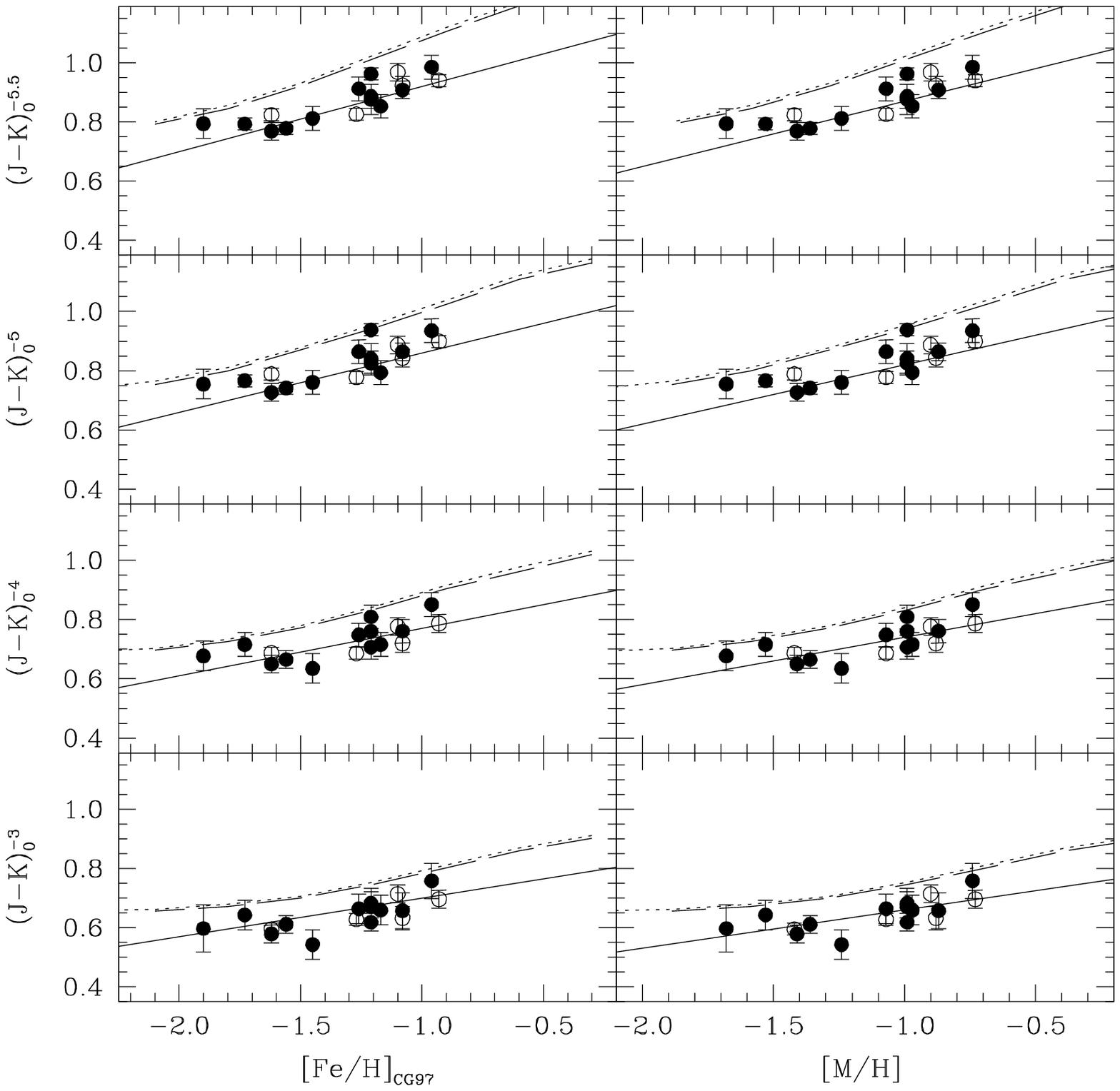}
      \caption{RGB $(J-K)_o$ color indices at fixed magnitudes of $M_K$ as a function
               of the metallicity [Fe/H]$_{CG97}$ and the global metallicity [M/H].
               Filled circles and open circles represent
               12 metal-poor bulge clusters observed here and
               5 bulge clusters in ~\citet{Kim06}, respectively.
               Solid lines are the calibration relations of~\citet{Val04a}.
               Dotted and dashed lines are the theoretical predictions of
               the Yonsei-Yale isochrones ~\citep{Kim02,Yi03} at $t = 12$ Gyr and 10 Gyr, respectively.
               \label{Fig3}}
   \end{figure}

   \begin{figure}
   \centering
   \includegraphics[width=0.45\textwidth]{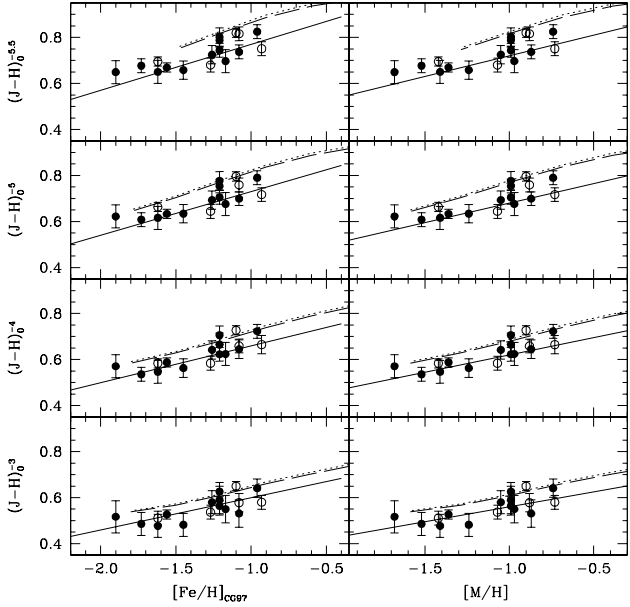}
      \caption{RGB $(J-H)_o$ color indices at fixed magnitudes of $M_H$ as a function
               of the metallicity [Fe/H]$_{CG97}$ and the global metallicity [M/H].
               Symbols are the same as Figure~\ref{Fig3}.\label{Fig4}}
   \end{figure}

In Figure~\ref{Fig3} and Figure~\ref{Fig4}, we present the colors at fixed 
magnitudes of $M_K=M_H=(-5.5, -5, -4,$ and $-3)$ as functions of cluster metallicity [
Fe/H]$_{CG97}$ and global metallicity [M/H] for 17 programme clusters.
Filled and open circles represent samples of clusters in this paper and~\citet{Kim06}, respectively.
Solid lines are the calibration relations of~\citet{Val04a}
for globular clusters spanning a metallicity range of $-2.12 \leqslant$ [Fe/H] $\leqslant-0.49$ 
in the Galactic bulge and halo.
As can be seen in Figure~\ref{Fig3} and Figure~\ref{Fig4},
the trends of the RGB color indices of $(J-K)_o$ and $(J-H)_o$ as a function of metallicity
agree well with the calibrations put forward by~\citet{Val04a}.
The RGB color indices of $(J-K)_o$ and $(J-H)_o$ linearly scale with the cluster metallicity
as the RGB color indices are bluer for the metal-poor clusters than metal-rich clusters.
In addition the fit slope increases progressively toward the RGB tip.
Theoretical predictions of the RGB location parameters were extracted from the
Yonsei-Yale isochrones~\citep{Kim02,Yi03} in order to compare with
the observed relations of the RGB colors and cluster metallicity.
The dotted and dashed lines are the theoretically estimated $(J-K)_o$ and $(J-H)_o$ values
of the RGB location as a function of metallicity at $t=12$ Gyr and 10 Gyr, respectively.
While the overall trends of the theoretical models show a good correlation
with the observed data, it appears that there are systematic shifts
in the RGB colors in our results from the relations inferred from the Yonsei-Yale isochrones.
Indeed, the theoretical model colors of the RGB at [Fe/H]$=-1.5$ seem to be 
$\sim0.06-0.11$ mag redder in $(J-K)$ and $\sim0.04-0.09$  mag redder in $(J-H)$ 
than the empirical results of ~\citet{Val04a}.
In addition, the shifts of the model colors become larger toward the RGB tip.
The shifts can be understood as a combination of the uncertainties involved in the color calibration of
$log T$ into $(J-K)_o$ colors (Kim 2009, private communications) 
and in the magnitude transformation of $K$ into 2MASS system,
the errors in the abundance determinations, and the photometric errors
in the observed colors.

   \begin{figure}
   \centering
   \includegraphics[width=0.45\textwidth]{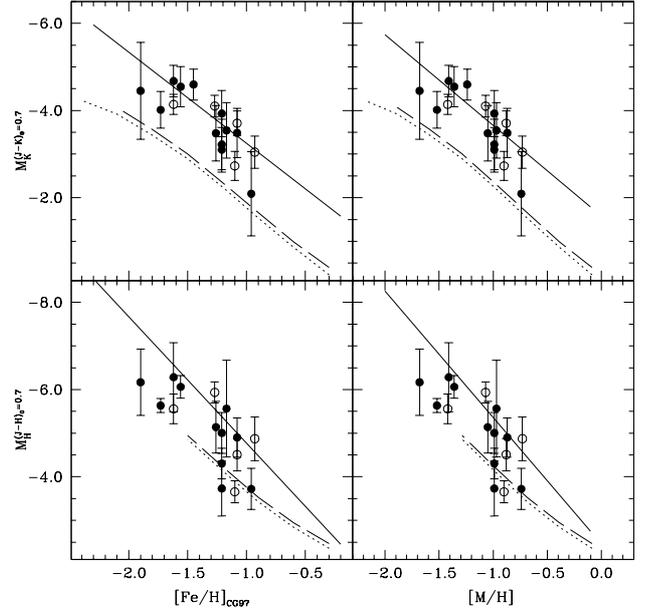}
      \caption{RGB magnitude indices $M_K$ and $M_H$ at fixed color $(J-K)_o=(J-H)_o=0.7$
               as a function of the metallicity [Fe/H]$_{CG97}$ and the global metallicity [M/H].
               Symbols are the same as Figure~\ref{Fig3}.\label{Fig5}}
   \end{figure}

Figure~\ref{Fig5} shows the dependence of the absolute magnitudes of $M_K$ and $M_H$
at fixed colors of $(J-K)_o=(J-H)_o=0.7$ on metallicity of 
the clusters investigated in this study.
The values measured in our sample fit well with the calibration relations (solid lines) of~\citet{Val04a}.
We note, however, the observed clusters in this paper show a larger scattered distribution of
the $M_K$ and $M_H$ magnitudes, compared with the distribution of metal-rich bulge globular clusters
and halo clusters of~\citet{Val04a}. 
This is possibly due to the uncertainty
in the derived absolute magnitudes associated with errors in the distance and reddening,
and errors in the polynomial fitting measurements on the fiducial normal points.
In fact,~\citet{Val04a} noted that errors in color of a few hundredths of a magnitude produce uncertainties of
about $0.2-0.3$ in $K$ magnitude, depending on the RGB region intercepted.
On the other hand, theoretical predictions of the absolute magnitudes of $M_K$ and $M_H$
at fixed colors of $(J-K)_o=(J-H)_o=0.7$ from Yonsei-Yale isochrones (dotted and dashed lines)
seem to be much fainter than the observed calibrations.
The magnitude shifts of the theoretical models at [Fe/H]$=-1.5$
are $\sim1.3-1.4$ mag fainter in $M_K$ and $M_H$ 
than the empirical results of ~\citet{Val04a}.
Here, we attribute the discrepancy to the uncertainties 
in the color calibration of the theoretical isochrone models.

   \begin{figure}
   \centering
   \includegraphics[width=0.45\textwidth]{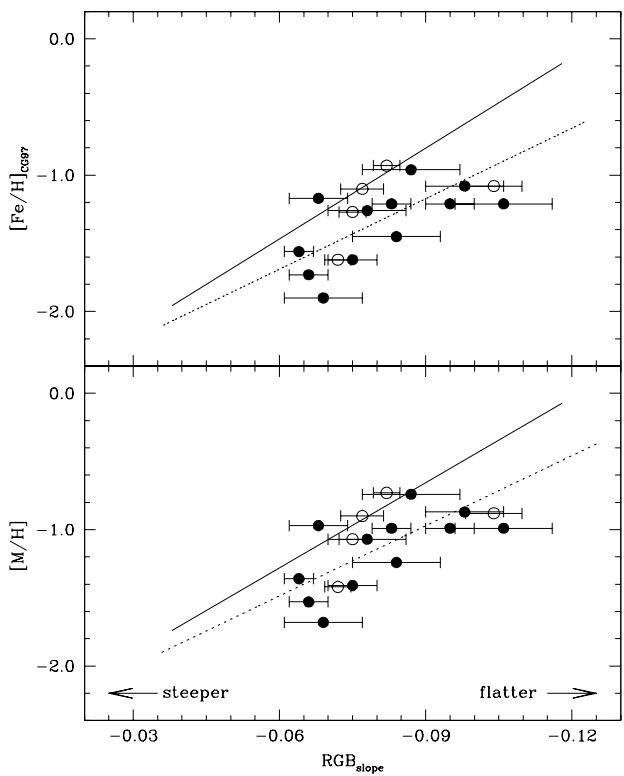}
      \caption{The RGB slope as a function of metallicity.
               The dotted lines are the relations found by~\citet{Iva02}.
               The other symbols are the same as Figure~\ref{Fig3}.\label{Fig6}}
   \end{figure}

The RGB slope is a useful parameter as it provides a photometric estimate of cluster metallicity.
Indeed, the RGB slope becomes progressively flatter with increasing metallicity, mainly
because the enhanced molecular blanketing could result in redder colors at a constant temperature
in the coolest and brightest stars~\citep{Ort91,Kuc95}. Moreover,
the RGB slope in a CMD is independent of reddening and the distance of a cluster.
Figure~\ref{Fig6} shows the measured RGB slopes as a function of metallicity
for the 17 programme clusters
with the empirical calibration relations (solid lines) from~\citet{Val04a}
and the theoretical predictions (dotted lines) from~\citet{Iva02}.
It is apparent in Figure~\ref{Fig6} that the trends for the dependency of the RGB slopes
on the metallicity is consistent with previous observational calibrations and
theoretical predictions, i.e., the steeper the RGB slope, the lower the metallicity of the cluster.
We also find a good consistency in the theoretical predictions of the distribution of
the observed RGB slopes for the metal-poor bulge globular clusters.
However, the estimated values of the RGB slopes for the observed metal-poor clusters
tend to be flatter at a given metallicity than the corresponding values 
in the previous empirical calibrations of ~\citet{Val04a}
for the metal-rich bulge clusters and the halo clusters.
This disagreement between our results and the relations found by ~\citet{Val04a}
is presumably due to the difference in the methods used to determine the RGB slope.
Indeed, ~\citet{Val04a} fit the fiducial ridge line of the RGB
in a magnitude range between 0.5 and 5 mag fainter than the brightest stars of each cluster,
while we used the fiducial normal points in a magnitude range between 0 and 5 mag fainter 
than the brightest point to keep consistency with the results in ~\citet{Kim06}.
We also note that the discrepancy might stem from the difficulty in estimating the RGB slope
for the metal-poor globular clusters in $(J-K,K)$ plane, especially
where the RGB is steeper than in any other plane, as mentioned in~\citet{Val04a}.
In particular, the near-infrared CMDs in the magnitude range that fits the RGB slope
for a metal-poor globular cluster can be contaminated by HB stars
despite statistical decontaminations from the RGB stars.
This is because the HB is not horizontal at all but slanted close to the RGB
in the near-infrared CMDs.

   \begin{figure*}
   \centering
   \includegraphics[width=0.93\textwidth]{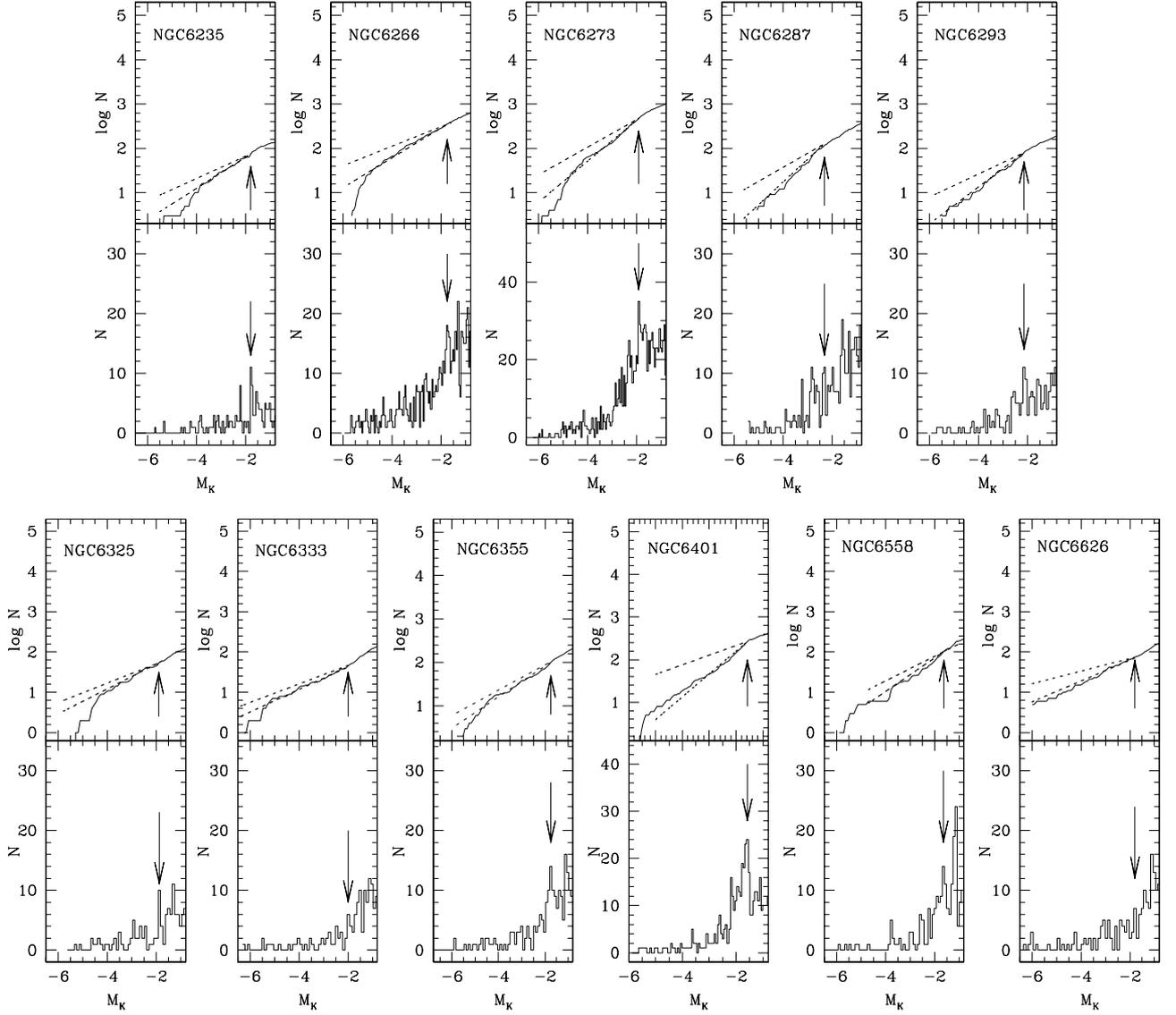}
      \caption{The logarithmic cumulative ({\it upper}) and differential ({\it lower}) LFs for RGB stars in the observed clusters.
               The arrows indicate the RGB bump position. The dashed lines in the cumulative LF
               are the linear fit to the regions above and below the RGB bump.\label{Fig7}}
   \end{figure*}

\begin{table*}
\caption{Magnitudes of the RGB bump and tip for the observed bulge clusters.}
\label{tbl5}
\centering
\begin{tabular}{c c c c c c c}
\hline\hline
Target & $K^{Bump}$ & $M_K^{Bump}$ & $M_{bol}^{Bump}$ & $K^{Tip}$ &
$M_K^{Tip}$ & $M_{bol}^{Tip}$ \\
\hline
\multicolumn{7}{c}{This paper} \\
\hline
NGC 6235    &13.39$\pm$0.06&  -1.79$\pm$0.06&   0.06$\pm$0.21& ... & ... & ...  \\
NGC 6266    &12.67$\pm$0.05&  -1.76$\pm$0.05&   0.08$\pm$0.21& ... & ... & ...  \\
NGC 6273    &12.82$\pm$0.05&  -1.91$\pm$0.05&  -0.28$\pm$0.21& 8.78$\pm$0.05&  -5.96$\pm$0.05& -3.43$\pm$0.21 \\
NGC 6287    &(13.56$\pm$0.07)&  (-2.27$\pm$0.07)&  (-0.33$\pm$0.21)& 9.88$\pm$0.07&  -5.96$\pm$0.07&  -3.32$\pm$0.21 \\
NGC 6293    &12.80$\pm$0.08&  -2.14$\pm$0.08&  -0.30$\pm$0.22& 9.28$\pm$0.08&  -5.66$\pm$0.08&  -3.25$\pm$0.22 \\
NGC 6325    &12.55$\pm$0.10&  -1.88$\pm$0.10&   0.1$\pm$0.22& ... & ... & ... \\
NGC 6333    &12.83$\pm$0.10&  -1.98$\pm$0.10&  -0.09$\pm$0.22& 8.82$\pm$0.10&  -5.99$\pm$0.10&  -3.53$\pm$0.22 \\
NGC 6355    &13.14$\pm$0.09&  -1.74$\pm$0.09&   0.09$\pm$0.22& ... & ... & ...  \\
NGC 6401    &13.29$\pm$0.06&  -1.58$\pm$0.06&   0.55$\pm$0.21& ... & ... & ...  \\
NGC 6558    &12.83$\pm$0.09&  -1.63$\pm$0.09&   0.21$\pm$0.22& ... & ... & ...  \\
NGC 6626    &(11.96$\pm$0.10)&  (-1.79$\pm$0.10)&   (0.01$\pm$0.20)&  7.45$\pm$0.10&  -6.29$\pm$0.10&  -3.69$\pm$0.22 \\
Terzan 4    & ... & ... & ... &  10.06$\pm$0.03&  -5.80$\pm$0.03&  -3.38$\pm$0.20 \\
\hline
\multicolumn{7}{c}{~\citet{Kim06}} \\
\hline
NGC 6541    &11.74$\pm$0.09 & -2.36$\pm$0.09&   -0.48$\pm$0.22 & 8.59$\pm$0.09 & -5.67$\pm$0.09 & -3.22$\pm$0.22 \\
NGC 6642    &13.23$\pm$0.10 & -1.41$\pm$0.10&    0.30$\pm$0.22 & 8.29$\pm$0.10 & -6.35$\pm$0.10 & -3.53$\pm$0.22 \\
NGC 6681    &13.13$\pm$0.04 & -1.90$\pm$0.04&   -0.02$\pm$0.20 & 8.69$\pm$0.04 & -6.34$\pm$0.04 & -3.78$\pm$0.20 \\
NGC 6717    &12.98$\pm$0.05 & -1.53$\pm$0.05&    0.57$\pm$0.21 & 9.02$\pm$0.05 & -5.49$\pm$0.05 & -2.84$\pm$0.21 \\
NGC 6723    &13.19$\pm$0.05 & -1.54$\pm$0.05&    0.50$\pm$0.21 & 8.73$\pm$0.05 & -6.00$\pm$0.05 & -3.37$\pm$0.21 \\
\hline
\end{tabular}
\end{table*}

\subsection{The RGB Bump and Tip}

The RGB bump on the CMDs has a crucial astrophysical significance
for the post-main-sequence evolution of low-mass stars in a globular cluster.
The position of the RGB bump for a globular cluster
depends on the chemical composition, the age, and other parameters controlling the internal evolution of a star.
Theoretical models of stellar evolution~\citep[e.g.,][]{Tho67,Ibe68} predict that,
at some level in the hydrogen burning shell stage in the RGB after the first dredge-up in a star,
the innermost penetration of the convective envelope inside star generates a discontinuity in the
hydrogen distribution profile.
When the advancing hydrogen burning shell passes through the generated discontinuity,
a star is expected to experience an evolutionary hesitation revealed as a temporary drop in luminosity
and a change in the evolutionary rate along the RGB.
This yields the RGB bump on the CMDs of stars in a globular cluster.

The detection of the RGB bump has been the subject of many studies from an empirical 
point of view~\citep[e.g.,][]{Fusi90,Fer99,Cho02,Val04b,Val07,Kim06},
suggesting that the combined use of the differential and integrated luminosity functions (LFs) of the RGB stars
is the best way to properly detect the RGB bump.
We note, however, it is more difficult to detect the RGB bumps in
the metal-poor globular clusters than in the metal-rich ones,
because of the small number of stars along the bright part of the RGB sequence.
Indeed, as mentioned in~\citet{Val04b},
the RGB bumps for the metal-poor clusters occur in the brightest portion of the RGB,
which is poorly populated sequence because of the high evolutionary rate of stars at the very end of the RGB.
Using the near-infrared LFs of the RGB stars,
~\citet{Val07} recently determined the RGB bumps for
Galactic bulge globular clusters with metallicities in the range of $-1.73\leqslant$ [Fe/H] $\leqslant-0.17$,
and presented new calibrations of the relation
between the cluster metallicity and the brightness of the RGB bump in the $K$ and bolometric magnitudes,
which differ from those in~\citet{Val04b} only in the metal-rich ends.

   \begin{figure}
   \centering
   \includegraphics[width=0.45\textwidth]{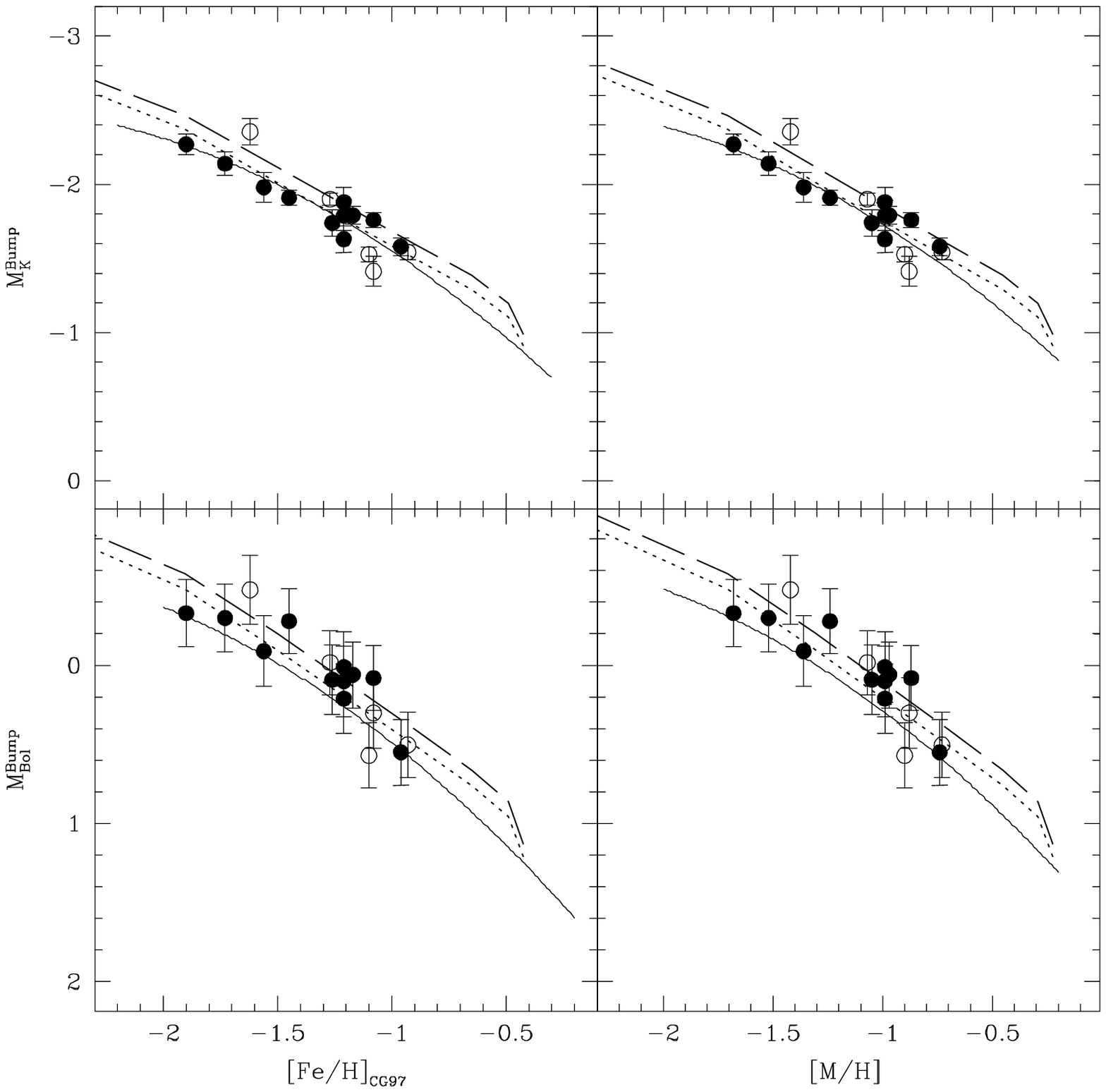}
      \caption{The behavior of the $M_K$ and $M_{bol}$ magnitudes
               of the RGB bumps for the observed clusters as a function of metallicity [Fe/H]$_{CG97}$
               and global metallicity [M/H].
               Symbols are the same as Figure~\ref{Fig3}.\label{Fig8}}
   \end{figure}

To construct the LF of the RGB stars for the observed globular clusters in this study,
we used the RGB stars selected to define the fiducial normal points of the $(J-K,K)$ CMDs for each cluster.
As mentioned in Sect. 3, the selected RGB stars used to estimate the fiducial normal points
include only RGB samples within $2\sigma$ deviation of the mean color for a given magnitude bin,
from which we properly avoided contaminations from other populations of stars,
such as asymptotic giant branch star, HB stars, and foreground field stars.
Considering the sample size of the RGB stars, we adjusted the size of the magnitude bins
of the LFs for each cluster, which enabled us to detect the RGB bump with an appropriate measurement error.
Figure~\ref{Fig7} shows the differential LF and the logarithmic cumulative LF of the RGB stars
for the observed 11 globular clusters.
We defined the RGB bump at a significant peak in the differential LF with a break in slope
of the logarithmic cumulative LF for the RGB stars in a cluster.
In the case of Terzan 4, the RGB bump could not be measured, as the RGB sample
is not sufficiently large to reach a safe detection of the bump. 
Magnitudes of RGB bumps for NGC 6287 and NGC 6626 are not clearly 
detected in the differential LFs. Instead, the clusters
show breaks in the slopes of the cumulative LFs
at the magnitudes of which we assign the RGB bumps for the clusters.
Applying the distance modulus and the reddening value for each cluster in Table~\ref{tbl3},
the determined $K$ magnitudes of the RGB bumps were transformed into the absolute $M_K$ magnitudes.
Then, the bolometric corrections for population II giant stars provided by~\citet{Mon98}
were used to convert the absolute magnitudes $M_K$ of the RGB bump into the bolometric magnitude $M_{bol}$.

In columns (2)-(4) of Table~\ref{tbl5}, we list the observed $K$, the absolute $M_K$,
and the bolometric $M_{bol}$ magnitudes of the RGB bumps for the observed clusters
in addition to those for 5 bulge clusters in~\citet{Kim06}.
The magnitude values of RGB bumps for NGC 6287 and NGC 6626,
which were determined from their cumulative LFs of the RGB stars,
are in parenthesis.
Errors in $K$ and $M_K$ are measurement errors, and those in $M_{bol}$ are
a combination of measurement errors and the global uncertainty of the distance moduli,
which is assumed to be 0.2 mag~\citep[e.g.,][]{Cho02}.
Figure~\ref{Fig8} plots the determined $M_K$ and $M_{bol}$ of the RGB bumps
versus cluster metallicity [Fe/H]$_{CG97}$ and global metallicity [M/H],
indicating that the RGB bump moves to fainter locations with increasing cluster metallicity.
As shown in Figure~\ref{Fig8}, the determinations of the RGB bumps for the metal-poor clusters
are consistent with the new calibrations for the Galactic bulge clusters (solid curves) of~\citet{Val07}.
The dotted and dashed lines indicate the theoretical predictions of the RGB bump magnitudes
as a function of metallicity from the Yonsei-Yale isochrones at $t=12$ Gyr and 10 Gyr~\citep{Kim02,Yi03},
showing a good agreement with the observations.

   \begin{figure}
   \centering
   \includegraphics[width=0.45\textwidth]{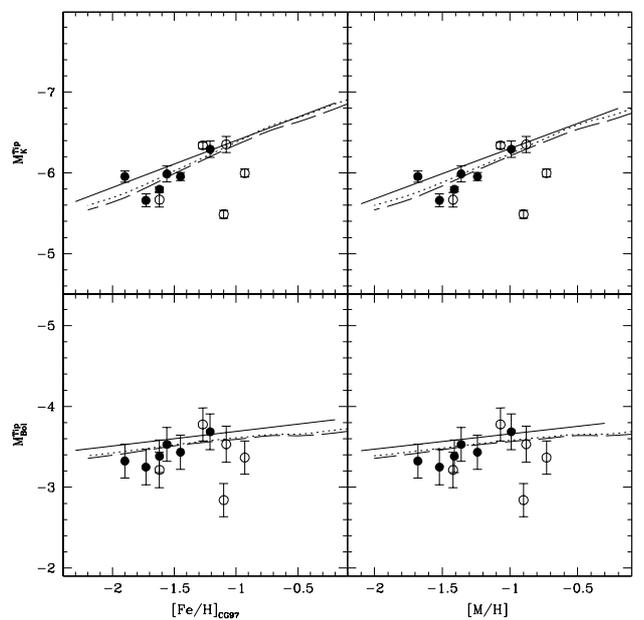}
      \caption{The behavior of the $M_K$ and $M_{bol}$ magnitudes
               of the RGB tip for the observed clusters as a function of metallicity [Fe/H]$_{CG97}$
               and global metallicity [M/H].
               Symbols are the same as Figure~\ref{Fig3}..\label{Fig9}}
   \end{figure}

The RGB tip (TRGB) is the evolution along the RGB ends with helium ignition in the stellar core.
Because the luminosity of the TRGB depends on the helium core mass which is fairly constant
over a large part of the low mass star range~\citep{Sal02},
the TRGB has a roughly constant brightness unrelated to the age of the population.
Thus, the luminosity of the TRGB is widely used as a standard candle to estimate the distance to
galaxies of any morphological type~\citep[e.g.,][]{Lee93,Mad95,Wal03}.
Recently, this method has also been carried out in near-infrared observations
to estimate the distances of nearby galaxies and Galactic globular clusters
~\citep[e.g.,][]{Mon95,Cio00,Cio05,Bel04}.
In this paper, the TRGB $K$ magnitude of the observed globular clusters
were determined from the brightness measurements of the brightest RGB stars
and the bright end of the observed LF of the RGB stars.
We note however, we were able to determine the $K$ magnitudes of the TRGBs only for 6 clusters
(NGC 6273, NGC 6287, NGC 6293, NGC 6333, NGC 6626, and Terzan 4), because
the brightest RGB is too poorly populated to define the TRGB
in the limited area of the other observed clusters.
Following the case of the RGB bumps, we estimated the absolute $M_K$
and the bolometric $M_{bol}$ magnitudes of the TRGB for the observed clusters.

The measured $K$, $M_K$, and $M_{bol}$ magnitudes of the TRGB are listed in the columns (5)-(7) of Table~\ref{tbl5}.
Similar to the $M_{bol}$ of the RGB bumps, errors in $M_{bol}$ for the TRGB are
a combination of measurement errors and the global uncertainty of 0.2 mag of the distance moduli.
Figure~\ref{Fig9} shows the relationship between the $M_K$ and $M_{bol}$ of the TRGB and the cluster metallicity
of the observed 6 clusters in addition to those of the 5 clusters in~\citet{Kim06},
indicating a good correlation with the previous calibrations (solid lines) of~\citet{Val04a}.
As noted in~\citet{Kim06}, the values of the TRGB for a compact post-core-collapse cluster NGC 6717
show a significant deviation from the calibration relations, because the number of bright RGB stars are still too small
to accurately measure the TRGB on the observed CMDs.
In Figure~\ref{Fig9} we overlay the theoretical predictions of the TRGB magnitudes
as a function of metallicity estimated from the Yonsei-Yale isochrones
~\citep{Kim02,Yi03}, which also seems to be consistent with the observations.

\section{Summary and Conclusions}

Detailed analyses of the RGB morphology for 12 metal-poor ([Fe/H] $\leq-1.0$) 
globular clusters in the Galactic bulge direction
have been performed using the high-quality near-infrared $JHK^{'}$ photometry.
From the study of the RGB shapes in the near-infrared CMDs for each cluster,
we measured photometric parameters, such as, the colors at different
magnitude levels, the magnitudes at different colors, and the RGB slopes.
The magnitudes of the RGB bump and tip, as major RGB evolutionary features, have also been determined
from the LFs of the selected RGB stars in each cluster.
The determined indices of the RGB morphology for the 12 observed clusters
have been combined with the results for 5 bulge clusters in~\citet{Kim06},
thus the entire dataset comprises $\sim75\%$ of the
total 22 metal-poor ([Fe/H] $\leq-1.0$) globular clusters
within 3 kpc from the Galactic center.
The behavior of the RGB morphology for the programme clusters has been
compared with the previous empirical calibration relations as a function of cluster metallicity
for the Galactic bulge globular clusters by~\citet{Val04a,Val07}
and theoretical predictions of the Yonsei-Yale isochrones~\citep{Kim02,Yi03}.
The results are summarized as follows:

\begin{enumerate}
\item Photometric indices for the RGB color at fixed magnitudes, $M_K=M_H=(-5.5, -5, -4,$ and $-3)$, and
      the RGB magnitudes at fixed colors, $(J-K)_o = (J-H)_o = 0.7$ have
      been measured from the fiducial normal points of the near-infrared $(J-K,K)$ and $(J-H,H)$ CMDs.
      Our results indicate that the correlations between the derived RGB indices and the cluster metallicity
      for the metal-poor globular clusters in the Galactic bulge direction are consistent with
      previous observational calibration relations for a sample of the metal-rich bulge clusters and the halo clusters~\citep{Val04a}.
      The trends of the theoretical models reliably represent the observed RGB color and magnitude indices,
      although there appears to be systematic shifts in color and magnitude, as a result of
      the uncertainties in the theoretical calculations and observational measurements.
\item The RGB slopes have been estimated from the determined fiducial normal points at the magnitude range
      between 0 and 5 magnitude fainter than the brightest point of the RGB.
      The distribution of the RGB slopes for the observed clusters
      show an expected evolutionary feature, i.e., the lower metallicity of the cluster, the steeper the RGB slope,
      while the RGB slopes for the programme clusters
      tend to be slightly flatter than those in the previous calibrations of~\citet{Val04a}.
\item The absolute $M_K$ and bolometric $M_{bol}$ magnitudes of the RGB bump and tip for the observed clusters
      have been determined from the differential and cumulative LFs of the selected RGB stars.
      The correlations between the cluster metallicity and the derived magnitudes of the RGB bump and tip for the
      metal-poor clusters in the Galactic bulge direction are consistent with the recent calibration relations
      for the Galactic bulge clusters~\citep{Val07}.
\end{enumerate}

Of a total of 17 metal-poor clusters presented in this paper,
only two clusters NGC 6266 and NGC 6723 have the cluster's orbital data~\citep{Din99,Din03},
indicating that NGC 6723 is a halo member passing the Galactic bulge at this moment and
NGC 6266 is associated with the motion of the Galactic thick disk.
Together with the derived RGB morphological properties,
further information about detailed orbital data
will provide more robust constraints on the role of the metal-poor globular clusters
in the formation of the Galactic bulge.

\begin{acknowledgements}
      This work has been supported by the Korea Research Foundation Grant funded by the Korea Government
      (KRF 2007-313-C00321), and also partly supported by Korea Astronomy and Space Science Institute
      (KASI 2009220000 and Yonsei-KASI Joint Research Program for Frontiers of Astronomy and Space Science), 
      for which we are grateful.
\end{acknowledgements}

\bibliographystyle{aa}
\bibliography{references}

\begin{thebibliography}{}
\bibitem[Bellazzini et al.(2004)]{Bel04}Bellazzini, M., Ferraro, F. R., Sollima, A., Pancino, E., \& Origlia, L.
	2004, \aap, 424, 199
\bibitem[Bonatto \& Bica(2008)]{Bon08}Bonatto, C., \& Bica, E. 2008, \aap, 479, 741
\bibitem[Carreta \& Gratton(1997)]{Carr97}Carreta, E., \& Gratton, R. G. 1997, A\&AS, 121, 95
\bibitem[Carpenter(2001)]{Carp01}Carpenter, J. M. 2001, \aj, 121, 2851
\bibitem[Carraro(2005)]{Car05}Carraro, G. 2005, \apj, 621, 61
\bibitem[Cho \& Lee(2002)]{Cho02}Cho, D.-H., \& Lee, S.-G. 2002, \aj, 124, 977
\bibitem[Cioni \& Habing(2005)]{Cio05}Cioni, M.-R. L., \&  Habing, H. J. 2005, \aap, 429, 837
\bibitem[Cioni et al.(2000)]{Cio00}Cioni, M.-R. L, van der Marel, R. P., Loup, C., \& Habing, H. J.
	2000, \aap, 359, 601
\bibitem[C{\^ o}t{\' e} et al.(2000)]{Cot00}C{\^ o}t{\' e}, P., Marzke, R. O., West, M. J.,\&  Minniti, D.
 	2000, \apj, 533, 869
\bibitem[Dinescu et al.(1999)]{Din99}Dinescu, D. I., Girard, T. M., \& van Altena, W. F.
    1999, \aj, 117, 1792
\bibitem[Dinescu et al.(2003)]{Din03}Dinescu, D. I., Girard T. M., van Altena W. F., \& Lopez C. E. 
   2003, \aj, 125,1373
\bibitem[Ferraro et al.(1999)]{Fer99}Ferraro, F. R., Messineo, M., Fusi Pecci, F., de Palo, M. A., 
	Straniero, O., Chieffi, A., \& Limongi, M. 1999, \aj, 118, 1738
\bibitem[Ferraro et al.(2000)]{Fer00}Ferraro, F. R., Montegriffo, P., Origlia, L., \& Fusi Pecci, F. 2000, \aj, 119, 1282
\bibitem[Froebrich et al.(2007)]{Fro07}Froebrich, D., Meusinger, H., \& Scholz, A. 2007, \mnras, 377, 54
\bibitem[Fusi Pecci et al.(1990)]{Fusi90}Fusi Pecci, F., Ferraro, F. R., Crocker, D. A., Rood, R. T., \& Buonanno, R.
	1990, \aap, 238, 95
\bibitem[Harris(1996)]{Har96}Harris, W. E. 1996, \aj, 112, 1487
\bibitem[Iben(1968)]{Ibe68}Iben, I. Jr. 1968, \nat, 220, 143
\bibitem[Ivanov \& Borissova(2002)]{Iva02}Ivanov, V. D., \& Borissova, J. 2002, \aap, 390, 937
\bibitem[Kim et al.(2002)]{Kim02}Kim, Y.-C., Demarque, P., Yi, S. K., \& Alexander, D. R. 2002, \apjs, 143, 499
\bibitem[Kim et al.(2006)]{Kim06}Kim, J.-W., Kang, A., Rhee, J., et al. 2006, \aap, 459, 499
\bibitem[Kobulnicky et al.(2005)]{Kob05}Kobulnicky, H. A., et al. 2005, \aj, 129, 239
\bibitem[Kuchinski et al.(1995)]{Kuc95}Kuchinski, L. E., Frogel, J. A., Terndrup, D. M., \& Persson, S. E.
	1995, \aj, 109, 1131
\bibitem[Lee et al.(1993)]{Lee93}Lee, M. G., Freedman, W. L., \& Madore, B. F. 1993, \apj, 417, 553
\bibitem[Lee et al.(1994)]{Lee94}Lee, Y.-W., Demarque, P., \& Zinn, R. 1994, \apj, 423, 248
\bibitem[Madore \& Freedman(1995)]{Mad95}Madore, B. F., \& Freedman, W. L. 1995, \aj, 109, 1645
\bibitem[McWilliam \& Rich(1994)]{Mcw94}McWilliam, A., \& Rich, R. M. 1994, \apjs, 91, 749
\bibitem[Minniti \& Zoccali(2008)]{Min08}Minniti, D., \& Zoccali, M. 2008, IAUS, 245, 323
\bibitem[Montegriffo et al.(1995)]{Mon95}Montegriffo, P., Ferraro, F. R., Fusi Pecci, F., \& Origlia, L. 1995, \mnras, 276, 739
\bibitem[Montegriffo et al.(1998)]{Mon98}Montegriffo, P., Ferraro, F. R. Origlia, L., \& Fusi Pecci, F. 1998, \mnras, 297, 872
\bibitem[Nakasato \& Nomoto(2003)]{Nak03}Nakasato, N., \& Nomoto, K. 2003, \apj, 588, 842
\bibitem[Origlia et al.(2005)]{Ori05}Origlia, L., Valenti, E., Rich, R. M., \& Ferraro, F. R. 2005, \mnras, 363, 897
\bibitem[Ortolani(1999)]{Ort99}Ortolani, S. 1999, \apss, 265,355
\bibitem[Ortolani et al.(1991)]{Ort91}Ortolani, S., Barbuy, B., \& Bica, E. 1991, \aap, 249, 31
\bibitem[Ortolani et al.(1997)]{Ort97}Ortolani, S., Barbuy, B., \& Bica, E. 1997, \aap, 319, 850
\bibitem[Piotto et al.(2002)]{Pio02}Piotto, G., King, I. R., Djorgovski, S. G., et al. 2002, \aap, 391, 945
\bibitem[Rich et al.(1998)]{Rich98}Rich, R. M., Ortolani, S., \& Barbuy, B. 1998, \aj, 116, 1295
\bibitem[Rieke \& Lebofsky(1985)]{Rie85}Rieke, G. H. \& Lebofsky, M. J. 1985, \apj, 288, 618
\bibitem[Salaris et al.(1993)]{Sal93}Salaris, M., Chieffi, A., \& Straniero, O. 1993, \apj, 414, 580
\bibitem[Salaris et al.(2002)]{Sal02}Salaris, M., Cassisi, S., \& Weiss, A. 2002, \pasp, 114, 375
\bibitem[Schlegel et al.(1998)]{Sch98}Schlegel, D. J., Finkbeiner, D. P., \& Davis, M. 1998, \apj, 500, 525
\bibitem[Stephens \& Frogel(2004)]{Ste04}Stephens, A. W., \& Frogel, J. A. 2004, \aj, 127, 925
\bibitem[Stetson(1987)]{Stet87}Stetson, P. B. 1987, \pasp, 99, 191
\bibitem[Stetson \& Harris(1988)]{Stet88}Stetson, P. B., \& Harris, W. E. 1988, \aj, 96, 909
\bibitem[Thomas(1967)]{Tho67}Thomas, H.-C. 1967, Z.Astrophys, 67, 420
\bibitem[Valenti et al.(2004a)]{Val04a}Valenti, E., Ferraro, F. R., \& Origlia, L. 2004a, \mnras, 351, 1204
\bibitem[Valenti et al.(2004b)]{Val04b}Valenti, E., Ferraro, F. R., \& Origlia, L. 2004b, \mnras, 354, 815
\bibitem[Valenti et al.(2004c)]{Val04c}Valenti, E., Ferraro, F. R., Perina, S., \& Origlia, L. 2004c, \aap, 419, 139
\bibitem[Valenti et al.(2005)]{Val05}Valenti, E., Origlia, L. \& Ferraro, F. R. 2005,  \mnras, 361, 272
\bibitem[Valenti et al.(2007)]{Val07}Valenti, E., Ferraro, F. R., \& Origlia, L. 2007, \aj, 133, 1287
\bibitem[Valenti et al.(2010)]{Val10}Valenti, E., Ferraro, F. R., \& Origlia, L. 2010, \mnras, 402, 1729
\bibitem[Walker(2003)]{Wal03}Walker, A. R. 2003, in Stellar candles for the extragalactic distance scale, ed. D. 
	Alloin, \& W. Gieren (Springer), Lect. Notes Phys., 635, 265
\bibitem[Willman et al.(2005)]{Wil05}Willman, B., et al. 2005, \aj, 129, 2692
\bibitem[Yi et al.(1997)]{Yi97}Yi, S., Demarque, P., \& Kim, Y. -C. 1997, \apj, 482, 677
\bibitem[Yi et al.(2003)]{Yi03}Yi, S. K., Kim, Y. -C., \& Demarque, P. 2003, \apjs, 144, 259
\bibitem[Zinn(1985)]{Zin85}Zinn, R. J. 1985, \apj, 293, 424
\bibitem[Zoccali et al.(2003)]{Zoc03}Zoccali, M. et al. 2003, \aap, 399, 931
\end{thebibliography}

\end{document}